\documentclass[showpacs,twocolumn]{revtex4}

\usepackage{graphicx}

\def\crea{^{\dagger}}
\def\anni{^{\phantom{\dagger}}}

\newcommand{\rd}{\mathrm{d}}
\newcommand{\re}{\mathrm{e}}
\newcommand{\ri}{\mathrm{i}}
\newcommand{\Imm}{\mathrm{Im}\,}
\newcommand{\Ree}{\mathrm{Re}\,}
\newcommand{\const}{\mathrm{const}}
\newcommand{\reff}{\mathrm{eff}}
\newcommand{\fig}{Fig.~}

\newcommand{\rl}{\langle\langle}
\newcommand{\rr}{\rangle\rangle}

\begin{document}

\title{Mott transition in the asymmetric Hubbard model at half-filling within dynamical mean-field theory}

\author{Ihor V. Stasyuk}
\author{Orest B. Hera}
\email{hera@icmp.lviv.ua}
\affiliation{%
Institute for Condensed Matter Physics of the National Academy of
Sciences of Ukraine,\\ 1~Svientsitskii Str., 79011 Lviv, Ukraine}

\date{April 12, 2005}

\begin{abstract}
We apply the approximate analytic methods to the investigation of
the band structure of the asymmetric Hubbard model where the
chemical potentials and electron transfer parameters depend on the
electron spin (type of quasiparticles). The Hubbard-I and
alloy-analogy approximations are the simplest approximations which
are used. Within the alloy-analogy approximation, the energy band
of particles does not depend on the transfer parameter of
particles of another sort. It means that the gap in the spectrum
opens at the critical value $U_{\mathrm{c}}$ that is the same in
two different limiting cases: the Falicov-Kimball model and the
standard Hubbard model. The approximate analytic scheme of the
dynamical mean-field theory is developed to include into the
theory the scattering of particles responsible for the additional
mechanism (due to the transfer of particles of another sort) of
the band formation. We use the so-called GH3 approach that is a
generalization of the Hubbard-III approximation. The approach
describes the continuous Mott transition with the $U_{\mathrm{c}}$
value dependent on a ratio of transfer parameters of different
particles.
\end{abstract}



\maketitle

\section{Introduction}

The asymmetric Hubbard model is considered as a generalization of
the Falicov-Kimball model \cite{Fal69} and the Hubbard model
\cite{Hub63}. This model describes a system with two sorts of
mobile particles (ions, electrons or quasiparticles) with
different hopping integrals and different values of chemical
potentials.

The hopping of particles is described by creation and annihilation
operators and transfer parameters $t_{ij}^{\sigma}$. The
Hamiltonian of the model is
\begin{equation}
H=\sum_{i} H_i + \sum_{ij\sigma} t_{ij}^{\sigma}
a_{i\sigma}\crea a_{j\sigma}\anni\,,%
\label{hamiltonian}
\end{equation}%
where the singe-site part
\begin{equation}
H_i=-\sum_{\sigma} \mu_{\sigma} n_{i\sigma} + U
n_{i\uparrow}n_{i\downarrow}%
\label{hamiltonian_i}
\end{equation}%
includes chemical potentials $\mu_\sigma$ and a local on-site
repulsion $U$ ($n_{i\sigma}=a_{i\sigma}\crea a_{i\sigma}\anni$).

Due to a complexity of the problem the asymmetric Hubbard model is
much less investigated than its limiting cases: the
Falicov-Kimball and Hubbard models. The model was investigated in
the large $U$ limit at half-filling. The effective Hamiltonian
corresponding to the anisotropic Heisenberg model was derived
\cite{Koc94,Koc95} and the effective antiferromagnetic interaction
was discussed \cite{Gru00}. A phase separation phenomenon was
considered in the ground state of the asymmetric Hubbard model
\cite{Uel04}. Some progress has been achieved in the case of
one-dimensional chains \cite{Fat95,Fuk98,Mac02}.

The essential achievement in the theory of the strongly correlated
electron systems is connected with the development of the
dynamical mean-field theory (DMFT) proposed by Metzner and
Vollhardt \cite{Met89} for the Hubbard model (see also
Ref.~\onlinecite{Geo96} for a review). This method is based on the
assumption that the self-energy is a pure local (a single-site)
quantity
$\Sigma_{ij}^{\sigma}(\omega)=\Sigma^{\sigma}(\omega)\delta_{ij}$
and it becomes exact in the limit of infinite dimensions
\cite{Mul89}. The current growing interest to DMFT is related to
the new developing technique combining DMFT and the density
functional theory within the local density approximation (LDA).
The new approach LDA+DMFT \cite{Ani97} allows one to calculate the
electronic structure of real materials correctly taking into
account the strong local correlations (see
Refs.~\onlinecite{Hel03,Geo04} for reviews).

The Falicov-Kimball and Hubbard models were intensively
investigated within DMFT (see reviews
Refs.~\onlinecite{Geo96,Fre03}). In the Falicov-Kimball model,
particle densities of states can be calculated exactly
\cite{Bra89,Bra90,Bra91,Bra92,Zla01,Fre05}. However, the spectrum
of localized particles was investigated mostly at half-filling. In
the case of the half-filled Hubbard model, a number of numerical
and analytic approximate methods was developed and applied to the
investigation of metal-insulator transitions
\cite{Noa99,Kra00,Sch99,Bul99,Roz99,Bul00a,Ono01,Bul01,Eas03,Vol05,Kot05}.

We use the approximate analytic approach to the investigation of
the band structure of the asymmetric Hubbard model in DMFT
\cite{Sta03}. This method was developed for the Hubbard model
\cite{Sta00,Sta02} and it is based on a mapping of the problem
onto the effective single-site Hamiltonian with the auxiliary
Fermi-operators describing the environment of a given site. The
approach is based on the equations of motion for Hubbard operators
followed by the different time decoupling of the higher order
Green's functions. The irreducible parts are separated off using
projecting on the basis of fermionic Hubbard operators. This
approach gives DMFT equations in the approximation that is a
generalization of the Hubbard-III approximation and includes as
simple specific cases the modified alloy-analogy approximation
\cite{Her96} and the Hub\-bard-III approximation \cite{Hub64b}. We
call it the generalized Hubbard-III (GH3) approximation. Recently,
an alternative decoupling scheme in an equation of motion approach
to the Hubbard model in infinite dimensions has been proposed in
Ref.~\onlinecite{Jes05}.

This paper is organized as follows. In
Section~\ref{dmft_formalism}, we review the formalism of DMFT,
where particle hopping can be introduced in two different ways:
using the coherent potential or the effective single-site
Hamiltonian with the auxiliary Fermi-filed \cite{Sta00}. The
equation of motion approach with the projecting technique and the
different-time decoupling scheme is described in
Section~\ref{GH3_approximation}. This approach gives the DMFT
equations within the generalized Hubbard-III approximation. In
Section~\ref{other_approximations}, simpler approximations to the
asymmetric Hubbard model are introduced. In
Section~\ref{analytic_at_transition}, some analytic properties
concerning continuous metal-insulator transitions are derived. Our
results are discussed in Section~\ref{result_discussion}, and
concluding remarks are given in Section~\ref{conclusions}.

\section{Dynamical mean field theory}
\label{dmft_formalism}

The particle Green's function in the Matsubara representation is
defined as follows
\begin{equation}
G_{ij}^{\sigma}(\tau-\tau') = \langle \mathcal{T}_{\tau}
a_{i\sigma}\crea (\tau) a_{j\sigma}\anni (\tau') \rangle \,,
\end{equation}%
\begin{equation}
G_{ij}^{\sigma}(\omega_n) = \int_0^{\beta} G_{ij}^{\sigma}(\tau)
\re^{-\ri \omega_n \tau} \rd \tau , \quad
\omega_n=\frac{\pi(2n+1)}{\beta} \,,
\end{equation}%
where $\mathcal{T}_\tau$ denotes imaginary time ordering, and
$\beta=1/T$ is the inverse temperature.

For the Hamiltonian (\ref{hamiltonian}), the particle Green's
function can be written as a solution of the Larkin equation
\cite{Shv03}:
\begin{equation}
G_{ij}^{\sigma}(\omega_n)=\Xi_{ij\sigma}(\omega_n) + \sum_{lm}
\Xi_{il\sigma}(\omega_n) t_{lm}^{\sigma}
G_{mj}^{\sigma}(\omega_n),
\end{equation}%
or in momentum space
\begin{equation}
G_{\sigma}(\omega_n, \mathbf{k})=\Xi_{\sigma}(\omega_n,
\mathbf{k}) + \Xi_{\sigma}(\omega_n, \mathbf{k})
t_{\mathbf{k}}^{\sigma} G_{\sigma}(\omega_n, \mathbf{k}),
\end{equation}%
where $\Xi_{\sigma}(\omega_n, \mathbf{k})$ is the total
irreducible part; it is connected to the Dyson self-energy by the
relation:
\begin{equation}
\Xi_{\sigma}^{-1}(\omega_n,
\mathbf{k})=\ri\omega_n+\mu_{\sigma}-\Sigma_{\sigma}(\omega_n,
\mathbf{k}).
\end{equation}%

In the limit of high lattice dimensions $d\rightarrow\infty$, the
irreducible part becomes a single-site quantity
\cite{Mul89,Met91}:
\begin{equation}
\Xi_{ij\sigma}(\omega_n) = \Xi_{\sigma} \delta_{ij},\quad
\Xi_{\sigma}(\omega_n,\mathbf{k})=\Xi_{\sigma}(\omega_n).
\end{equation}%
The function $\Xi_{\sigma}(\omega_n)$ or
$\Sigma_{\sigma}(\omega_n)$ is calculated using the auxiliary
single-site problem. This problem corresponds to the following
replacement
\begin{equation}
\re^{-\beta H} \rightarrow \re^{-\beta H_{\reff}} = \re^{-\beta
H_0} \mathcal{T}_\tau \exp \bigg[ - \int_0^{\beta} \rd\tau
H_{\mathrm{int}}(\tau) \bigg],%
\label{h_eff1}
\end{equation}%
\begin{equation}
H_{\mathrm{int}}(\tau) = \int_0^{\beta} \rd\tau'  \sum_{\sigma}
J_{\sigma}(\tau-\tau') a_{\sigma}\crea (\tau) a_{\sigma}\anni
(\tau'),
\end{equation}%
where
\begin{equation}
H_0=H_i,
\end{equation}%
and $J_\sigma(\tau-\tau')$ is the coherent potential. This
function describes the propagation of a particle in the
environment without going through the given site between moments
$\tau$ and $\tau'$. It is determined self-consistently from the
condition that the same irreducible part determines the lattice
single-site Green's function
\begin{equation}
G_\sigma(\omega_n)=G_{ii}^\sigma(\omega_n)=\frac{1}{N}\sum_\mathbf{k}
G^\sigma(\omega_n, \mathbf{k}),%
\label{sys3}
\end{equation}%
\begin{equation}
G_\sigma (\omega_n,
\mathbf{k})=\frac{1}{\Xi_\sigma^{-1}(\omega_n)-t^\sigma_\mathbf{k}}%
\label{sys1}
\end{equation}%
and the Green's function of the effective single-site problem
\begin{eqnarray}
G_\sigma(\omega_n)=\frac{1}{\Xi_\sigma^{-1}(\omega_n)-J_\sigma(\omega_n)}\,.
 \label{sys2}
\end{eqnarray}%

To investigate the Green's functions in the time representation
the analytic continuation from the imaginary to real axis
($\ri\omega_n \rightarrow \omega + \ri \varepsilon $) is
performed:
\begin{equation}
G_{\sigma}(\omega_n) \rightarrow G_{\sigma}(\omega) = 2\pi \rl
a_{\sigma}\anni | a_{\sigma}\crea \rr_{\omega}\,.
\end{equation}%

The self-consistency condition is rewritten in the following form
\begin{eqnarray}
&&{}%
G_{\sigma}^{-1}(\omega)=\Xi_{\sigma}^{-1}(\omega)-J_{\sigma}(\omega),
\label{G_Xi_J}\\ &&{}%
G_{\sigma}(\omega)=\int_{-\infty}^{+\infty}
\frac{\rho^0_{\sigma}(t) \rd t}{\Xi_{\sigma}^{-1}(\omega)-t}\,,
\label{G_from_Xi}
\end{eqnarray}%
where (\ref{sys1}) is inserted into (\ref{sys3}) and the sum over
$\mathbf{k}$ is replaced by the integration with the
noninteracting density of states. The infinite dimensional
hypercubic lattice and the Bethe lattice are the most popular
lattices investigated in the dynamical mean-field theory. A
Gaussian noninteracting density of states corresponds to the
hypercubic lattice with nearest-neighbour hopping:
\begin{equation}
\rho_{\sigma}^{\mathrm{hyp}}(\varepsilon) =
\frac{1}{t_{\sigma}\sqrt{\pi}} \exp
\bigg(-\frac{\varepsilon^2}{t^{2}_{\sigma}} \bigg),
\end{equation}%
where the transfer parameters must be scaled to keep the kinetic
and interaction energies of the same order of magnitude in the
infinite dimensional limit \cite{Met89}:
\begin{equation}
 t^{\sigma}_{ij} = \frac{t_{\sigma}}{2\sqrt{d}}
\end{equation}%
when $i$ and $j$  are nearest-neighbours and zero otherwise. In
the case of the Bethe lattice with the coordination number
$z\rightarrow \infty$, the noninteracting density of states is
semielliptic
\begin{equation}
\rho_{\sigma}^{\mathrm{Bethe}}(\varepsilon) = \frac{1}{2 \pi
t^{2}_{\sigma}} \sqrt{4 t^{2}_{\sigma} - \varepsilon^2}, \quad
|\varepsilon|<2t_{\sigma} \,,
\end{equation}%
and the hopping integral for nearest-neighbours is scaled as
follows:
\begin{equation}
 t^{\sigma}_{ij}  = \frac{t_{\sigma}}{\sqrt{z}}\,.
\end{equation}%
Besides the hypercubic and Bethe lattices, the nonsymmetric
densities of states can be considered. The infinite dimensional
generalization of a face-centred-cubic lattice that was used to
describe ferromagnetic order in the Hubbard model \cite{Vol97} is
an example of such a lattice.

A solution of the single-site problem (\ref{h_eff1}) gives the
particle Green's function $G_{\sigma}(\omega)$ as a functional of
the coherent potential $J_{\sigma}(\omega)$. This dependence
supplemented by the self-consistency condition forms a closed set
of equations for the single-site Green's function and the coherent
potential.

We reformulate the single-site problem introducing the effective
Hamiltonian as it was done in Ref.~\onlinecite{Sta00}:
\begin{equation}
H_{\reff}=H_0+\sum_{\sigma} V_{\sigma}\big(a_{\sigma}\crea
\xi_{\sigma}\anni\! +\xi_{\sigma}\crea a_{\sigma}\anni
\!\big)+H_{\xi}.
\end{equation}
The auxiliary fermionic operators $\xi_{\sigma}\crea$,
$\xi_{\sigma}\anni$ are introduced to describe particle hopping
between the site ($H_0$) and effective environment defined by the
Hamiltonian $H_{\xi}$. The coherent potential is expressed as the
Green's function in unperturbed  Hamiltonian $H_{\xi}$:
\begin{equation}
J_{\sigma}(\omega)=2\pi V^2_{\sigma}
\langle\langle\xi_{\sigma}\anni | \xi_{\sigma}\crea
\rangle\rangle_{\omega}^{\xi}.
\end{equation}
Thus, an explicit form of $H_{\xi}$ is not required to solve the
problem.

\section{Generalized Hubbard-III approximation}
\label{GH3_approximation}

We use the single-site Hamiltonian of the asymmetric Hubbard model
written in terms of Hubbard operators
\begin{equation}
  H_{0}  =  - \sum_{\sigma} \big[  \mu_{\sigma} \big( X^{\sigma\sigma}+X^{22}
  \big) \big] + U X^{22}\,,
\end{equation}
acting on the basis of single-site states $| n_A, n_B\rangle$
\begin{equation}
\begin{array}{lll}
|0\rangle=|0,0\rangle , &  &\qquad |A\rangle=|1,0\rangle, \\
|2\rangle=|1,1\rangle, &  &\qquad |B\rangle=|0,1\rangle .
\end{array}
\label{BaseStates2}
\end{equation}
In this case, the particle creation and annihilation operators are
expressed as
\begin{equation}
a_{\sigma}= X^{0\sigma} + \zeta X^{\bar{\sigma} 2},
\end{equation}
and the two-time Green's function $G_\sigma(\omega) \equiv 2\pi
\langle\langle a_\sigma\anni | a_\sigma\crea
\rangle\rangle_\omega$ is expressed as:%
{\arraycolsep=2pt
\begin{eqnarray}
 G_{\sigma} & = & 2\pi\big[\langle\langle X^{0\sigma}| X^{\sigma 0}
\rangle\rangle_{\omega}
 +\zeta \langle\langle X^{0\sigma}| X^{2\bar{\sigma}}
 \rangle\rangle_{\omega}  \nonumber \\
& &{} +\zeta \langle\langle X^{\bar{\sigma}2}| X^{\sigma0}
\rangle\rangle_{\omega}
 +\langle\langle X^{\bar{\sigma}2}| X^{2\bar{\sigma}}
 \rangle\rangle_{\omega}\big]\,,
 \label{Grin_X}
\end{eqnarray}}%
where the following notations for sort indices are used:
$\bar{\sigma}=B$, $\zeta=+$ for $\sigma=A$ and $\bar{\sigma}=A$,
$\zeta=-$ for $\sigma=B$.

To calculate the Green's functions in (\ref{Grin_X}), we use the
equations of motion for Hubbard operators:
\begin{eqnarray}
\ri\frac{\rd}{\rd t} X^{0\sigma
(\bar{\sigma}2)}(t)=\big[X^{0\sigma (\bar{\sigma}2)}\,,
H_{\reff}\big]\,. \label{eq_motion01}
\end{eqnarray}%
The commutators (\ref{eq_motion01}) are projected on the subspace
formed by fermionic operators $X^{0\sigma}$ and
$X^{\bar{\sigma}2}$:
\begin{equation}
\big[X^{\gamma}, H_{\reff}\big]=\alpha_1^{\gamma} X^{0\sigma}+
\alpha_2^{\gamma} X^{\bar\sigma 2} + Z^{\gamma}\,.
\end{equation}
The operators $Z^{0\sigma (\bar{\sigma}2)}$ are defined as
orthogonal to the operators from the basic subspace
\cite{Sta00,Sta02,Sta03}:
\begin{equation}
\langle \{Z^{0\sigma (\bar{\sigma}2)}, X^{0\sigma
(\bar{\sigma}2)}\}\rangle = 0.%
\label{orthog}
\end{equation}
Thus, these equations (\ref{orthog}) determine the projecting
coefficients $\alpha_{i}^{0\sigma (\bar{\sigma}2)}$ which are
expressed in terms of mean values $\langle \xi_{\sigma}\anni
X^{pq} \rangle$ combined into the following constant
\begin{equation}
 \varphi_\sigma =
 \langle \xi_{\bar\sigma}\anni X^{\bar\sigma 0}\rangle +
 \zeta \langle X^{\sigma 2} \xi_{\bar\sigma}\crea \rangle
\end{equation}
and mean values of the Hubbard operators
\begin{equation}
A_{pq}= \langle X^{pp}+X^{qq}\rangle,\,\,
A_{0\sigma}=1-n_{\bar\sigma},\,\, A_{2\bar\sigma}=n_{\bar\sigma}.
\end{equation}

This procedure leads to the Green's functions of the $\rl
Z^{0\sigma (\bar{\sigma}2)}| X^{\sigma 0 (2 \bar{\sigma})} \rr$
type. The similar procedure can be applied with respect to the
second time argument. As a result, we come to the relations
between the components of the Green's function $G_\sigma$ and
scattering matrix $\hat{P}_\sigma$. In a matrix representation, we
have
\begin{equation}
 \hat{G}_\sigma=\hat{G}_0^\sigma+\hat{G}_0^\sigma\hat{P}_\sigma\hat{G}_0^\sigma,
\end{equation}
where
\begin{equation}
  \hat{G}_\sigma=2\pi \left(
  \begin{array}{lll}
  \langle\langle X^{0\sigma}| X^{\sigma 0} \rangle\rangle & &
  \langle\langle X^{0\sigma}| X^{2 \bar\sigma} \rangle\rangle \\
  \langle\langle X^{\bar\sigma 2}| X^{\sigma 0} \rangle\rangle & &
  \langle\langle X^{\bar\sigma 2}| X^{2 \bar\sigma} \rangle\rangle
  \end{array}
  \right),
  \label{Grin_Mart}
\end{equation}
and nonperturbed Green's function $\hat {G}^\sigma_0$ is
{\arraycolsep=1.6pt
\begin{equation}
  \hat{G}^\sigma_0=\frac{1}{D_\sigma} \left(
  \begin{array}{ccc}
  \omega-b_\sigma & & -\zeta\frac{V_\sigma}{A_{2\bar\sigma}}\varphi_\sigma \\
  -\zeta\frac{V_\sigma}{A_{0\sigma}}\varphi_\sigma & & \omega-a_\sigma
  \end{array}
  \right) \left(
  \begin{array}{ccc}
  A_{0\sigma} & & 0 \\
  0  & & A_{2\bar\sigma}
  \end{array}
  \right),
\end{equation}}
where
\begin{equation}
  D_\sigma=(\omega-a_\sigma)(\omega-b_\sigma)-
  \frac{V^2_\sigma}{A_{0\sigma}A_{2\bar\sigma}}\varphi_\sigma^2,
\end{equation}
\begin{equation}
  a_\sigma=-\mu_\sigma+\frac{V_\sigma}{A_{0\sigma}}\varphi_\sigma,
  \quad
  b_\sigma=U-\mu_\sigma+\frac{V_\sigma}{A_{2\bar\sigma}}\varphi_\sigma.
\end{equation}
The scattering matrix
{\arraycolsep=-5pt
\begin{eqnarray}
&&{}  \hat{P}_\sigma=2\pi
  \left(
  \arraycolsep=2pt
  \begin{array}{ccc}
  A_{0\sigma}^{-1} & & 0 \\
  0  & & A_{2\bar\sigma}^{-1}
  \end{array}
  \right) \nonumber\\
&&{}  \times \left(
  \arraycolsep=2pt
  \begin{array}{ccc}
  \langle\langle Z^{0\sigma}| Z^{\sigma 0} \rangle\rangle & &
  \langle\langle Z^{0\sigma}| Z^{2 \bar\sigma} \rangle\rangle \\
  \langle\langle Z^{\bar\sigma 2}| Z^{\sigma 0} \rangle\rangle & &
  \langle\langle Z^{\bar\sigma 2}| Z^{2 \bar\sigma} \rangle\rangle
  \end{array}
  \right)
  \left(
  \arraycolsep=2pt
  \begin{array}{ccc}
  A_{0\sigma}^{-1} & & 0 \\
  0  & & A_{2\bar\sigma}^{-1}
  \end{array}
  \right)
  \label{scattmatr}
\end{eqnarray}}%
contains the scattering corrections of the second and the higher
orders in powers of $V_\sigma$. The separation of the irreducible
parts in $\hat{P}_\sigma$ enables us to obtain the mass operator
$\hat{M}_\sigma=\hat{P}_\sigma|_{\mathrm{ir}}$ and the single-site
Green's function expressed as a solution of the Dyson equation
\begin{equation}
 \hat{G}_\sigma=(1-\hat{G}^\sigma_0\hat{M}_\sigma)^{-1} \hat{G}^\sigma_0.
 \label{dyson1}
\end{equation}

The mass operator $\hat {M}_\sigma$ is calculated in zero
approximation
\begin{equation}
 \hat{M}_{\sigma}= \hat{P}_{\sigma}^{(0)},
 \label{massop}
\end{equation}
where the time correlation functions related by the spectral
theorem to the irreducible Green's functions are calculated using
procedure of different-time decoupling. In our case it means an
independent averaging of the products of $X$ and $\xi$ operators.
For example,%
{\arraycolsep=-2pt
\begin{eqnarray}
&&{} \langle
  \xi_\sigma\crea(t)(X^{00}+X^{\sigma\sigma})_t
  (X^{00}+X^{\sigma\sigma})\xi_\sigma\anni
 \rangle^{\mathrm{ir}} \nonumber\\
&& {\qquad} \approx \langle (X^{00}+X^{\sigma\sigma})_t
 (X^{00}+X^{\sigma\sigma}) \rangle
 \langle
  \xi_\sigma\crea (t) \xi_\sigma\anni
 \rangle.
 \label{e43}
\end{eqnarray}}%

For simplicity we calculate the correlation functions for Hubbard
operators in zero approximation%
{\arraycolsep=-2pt
\begin{eqnarray}
&&{} \langle (X^{00}+X^{\sigma\sigma})_t
(X^{00}+X^{\sigma\sigma}) \rangle \nonumber\\
&&{\qquad}  \approx \langle (X^{00}+X^{\sigma\sigma})^2 \rangle =
A_{0\sigma}.
\end{eqnarray}}%
In this example, the following Green's function is reconstructed
from the correlation functions using the spectral representation
\cite{Sta00,Sta02,Sta03,Sta05}:
\begin{equation}
  I(\omega)=A_{0\sigma}\langle\langle \xi_\sigma\anni |
  \xi_\sigma\crea
  \rangle\rangle_\omega^\xi =
  \frac{A_{0\sigma}}{2\pi V^2} J_\sigma (\omega).
\end{equation}

Using the above procedure, we can obtain the final expressions for
the total irreducible part:
 {\arraycolsep=0.5pt
\begin{eqnarray}
 \Xi_\sigma^{-1}(\omega)&=&\bigg[ \frac{A_{0\sigma}}
       {\omega+\mu_\sigma-{\Omega}_{\sigma}(\omega)}
       + \frac{A_{2\bar\sigma}}
       {\omega+\mu_\sigma-U-{\Omega}_{\sigma}(\omega)}
       \bigg]^{-1} \nonumber\\
& &{}  + {\Omega}_{\sigma}(\omega),%
\label{irreduciblepart}
\end{eqnarray}}%
where
\begin{equation}
{\Omega}_{\sigma}(\omega)=J_\sigma(\omega)-
\frac{R_\sigma(\omega)}{A_{0\sigma}A_{2\bar\sigma}} +
\frac{V_\sigma \varphi_\sigma}{A_{0\sigma}A_{2\bar\sigma}}\,,%
\label{Omega}
\end{equation}
and {\arraycolsep=-4pt
\begin{eqnarray}
  & & {R_\sigma(\omega)  =  -\frac{\langle X^{\sigma\sigma}+
  X^{\bar\sigma\bar\sigma}\rangle}{2}
  J_{\bar\sigma}(\omega+\mu_\sigma-\mu_{\bar\sigma})}  \nonumber\\
  & & -\frac{\langle X^{\sigma\sigma}-X^{\bar\sigma\bar\sigma}\rangle}{2\pi}
  \!\int\limits_{-\infty}^{+\infty}\! \frac{\rd \omega' \Imm J_{\bar\sigma}(\omega'+\ri 0^+)}
  {\omega-\omega'-\mu_{\bar\sigma}+\mu_\sigma}
  \tanh\frac{\beta\omega'}{2} \nonumber\\
  & &+\frac{\langle X^{0 0}+ X^{2 2}\rangle}{2}
  J_{\bar\sigma}(U-\mu_\sigma-\mu_{\bar\sigma}-\omega) \nonumber \\
  & & {-}\frac{\langle X^{00}-X^{22}\rangle}{2\pi}
  \!\int\limits_{-\infty}^{+\infty}\!\! \frac{\rd \omega' \Imm  J_{\bar\sigma}(-\omega'-\ri 0^+)}
  {\omega-\omega'+\mu_{\bar\sigma}+\mu_\sigma-U}
  \tanh\frac{\beta\omega'}{2}\,.%
\label{R_omega}
\end{eqnarray}}%

The average values of the Hubbard operators are determined using
the spectral representation of corresponding Green's functions.
The particle density of states is calculated as an imaginary part
of the interacting single-particle Green's function:
\begin{equation}
\rho_\sigma(\omega)=-\frac{1}{\pi}\lim_{\eta \rightarrow 0^+} \Imm
G_\sigma(\omega+\ri\eta)%
\label{rho_DOS}
\end{equation}
giving the concentration
\begin{equation}
n_{\sigma}=\int_{-\infty}^{+\infty} \frac{\rho_\sigma(\omega)
\rd\omega}{\re^{\beta\omega}+1} \,.
\end{equation}
The Green's functions $\rl \xi_{\sigma}\anni|X^{pq}\rr_{\omega}$
that are required to find $\varphi_\sigma$ can be calculated from
the exact relation
\begin{equation}
 V_{\sigma} \rl \xi_{\sigma}\anni|X^{pq}\rr_{\omega}
 =  J_{\sigma}(\omega) \rl
a_{\sigma}|X^{pq}\rr_{\omega}
\end{equation}
derived in Ref.~\onlinecite{Sta05}. Thus, the approximation given
by the equations (\ref{irreduciblepart})--(\ref{R_omega}) we
called the GH3 approximation.

We consider below the special case at half-filling
($\mu_A=\mu_B=U/2$, $n_A=n_B=1/2$), and  due to the particle-hole
symmetry we have:
\begin{equation}
\varphi_\sigma=0,
\end{equation}%
\begin{equation}
\Omega_{\sigma}(\omega)=J_{\sigma}(\omega)+2
J_{\bar\sigma}(\omega),
\end{equation}%
and
\begin{equation}
\Xi^{-1}_{\sigma}(\omega)=\omega-\frac{U^2}{4[\omega -
J_{\sigma}(\omega) - 2 J_{\bar\sigma}(\omega) ]}\,.%
\label{Xi_halffilling}
\end{equation}%

For the standard Hubbard model ($t_A=t_B$, $J_A=J_B$) the
approximate solution of the single-site problem gives the usual
Hubbard-III approximation.

In the case of the Falicov-Kimball model ($t_B=0$,
$J_B(\omega)=0$) at half-filling, GH3 gives the following
expression:
\begin{equation}
G_B(\omega)=\frac{\omega-2J_A(\omega)}{\omega^2-U^2/4-2\omega
J_A(\omega)},%
\label{hf_FK}
\end{equation}%
where the coherent potential of itinerant particles $J_A$
corresponding  to the result of the alloy-analogy approximation is
calculated exactly.

\section{Other approximations to the asymmetric Hubbard model}
\label{other_approximations}

\subsection{Hubbard-I approximation}%

It was the Hubbard-I approximation \cite{Hub63} which in the
simplest way described band forming in the Hubbard model with the
strong on-site repulsion $U$. The total irreducible part in this
approximation reads
\begin{equation}
\Xi_{\sigma}(\omega)=\frac{1-n_{\bar\sigma}}{\omega+\mu_\sigma} +
\frac{n_{\bar\sigma}}{\omega+\mu_\sigma-U}\,.
\end{equation}%

\begin{figure}
\includegraphics[width=0.24\textwidth]{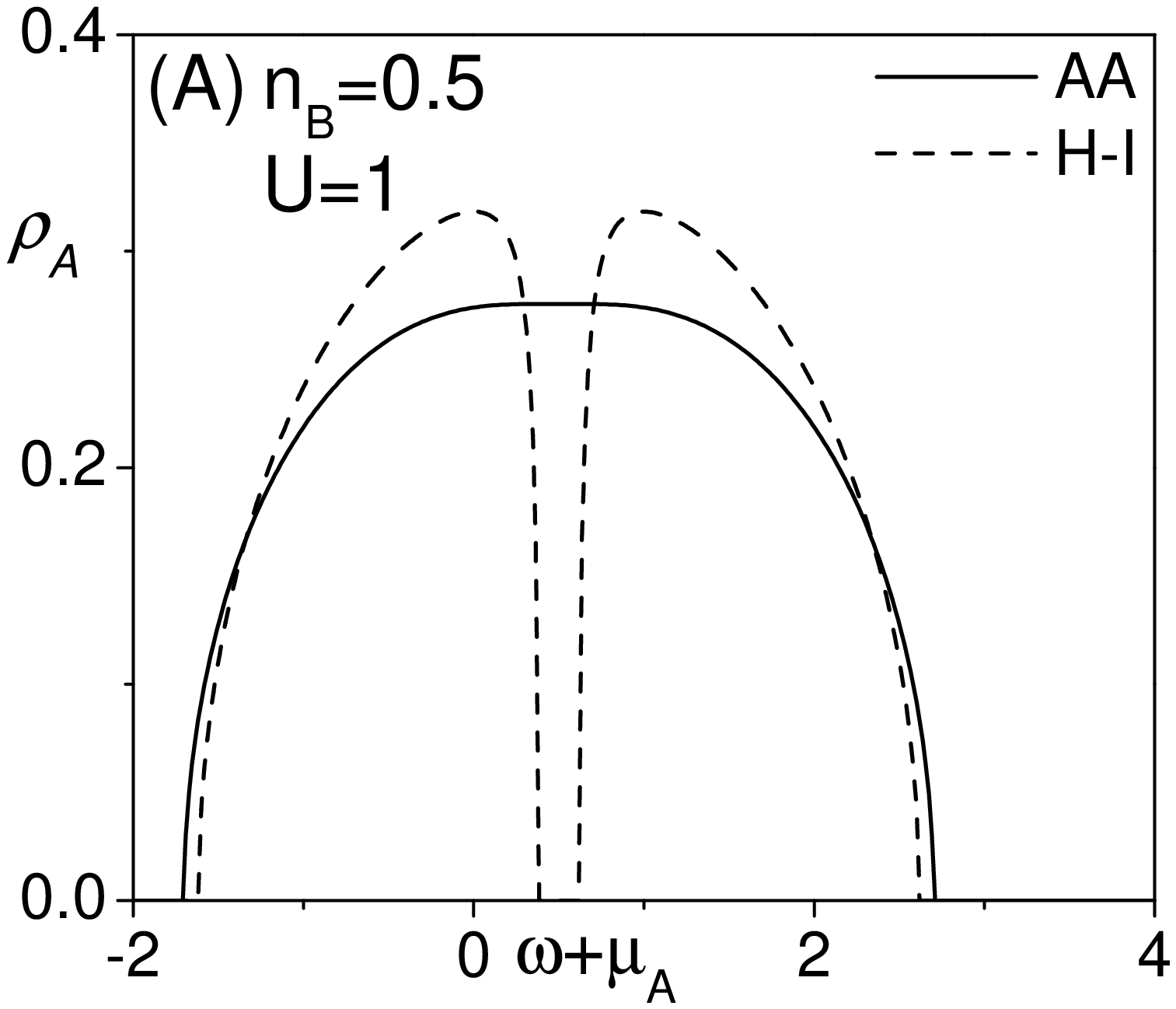}\hfill
\includegraphics[width=0.24\textwidth]{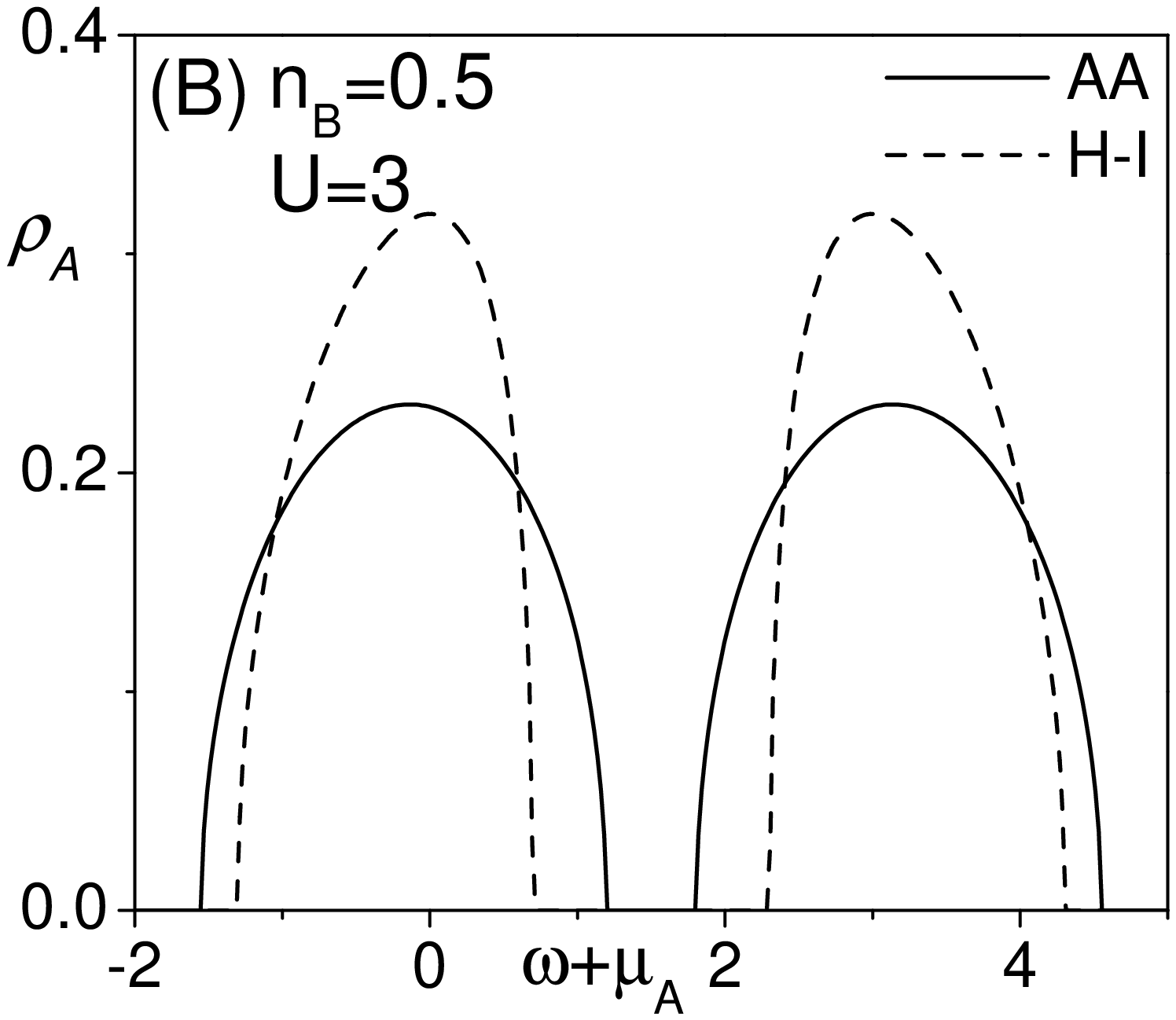}\\
\includegraphics[width=0.24\textwidth]{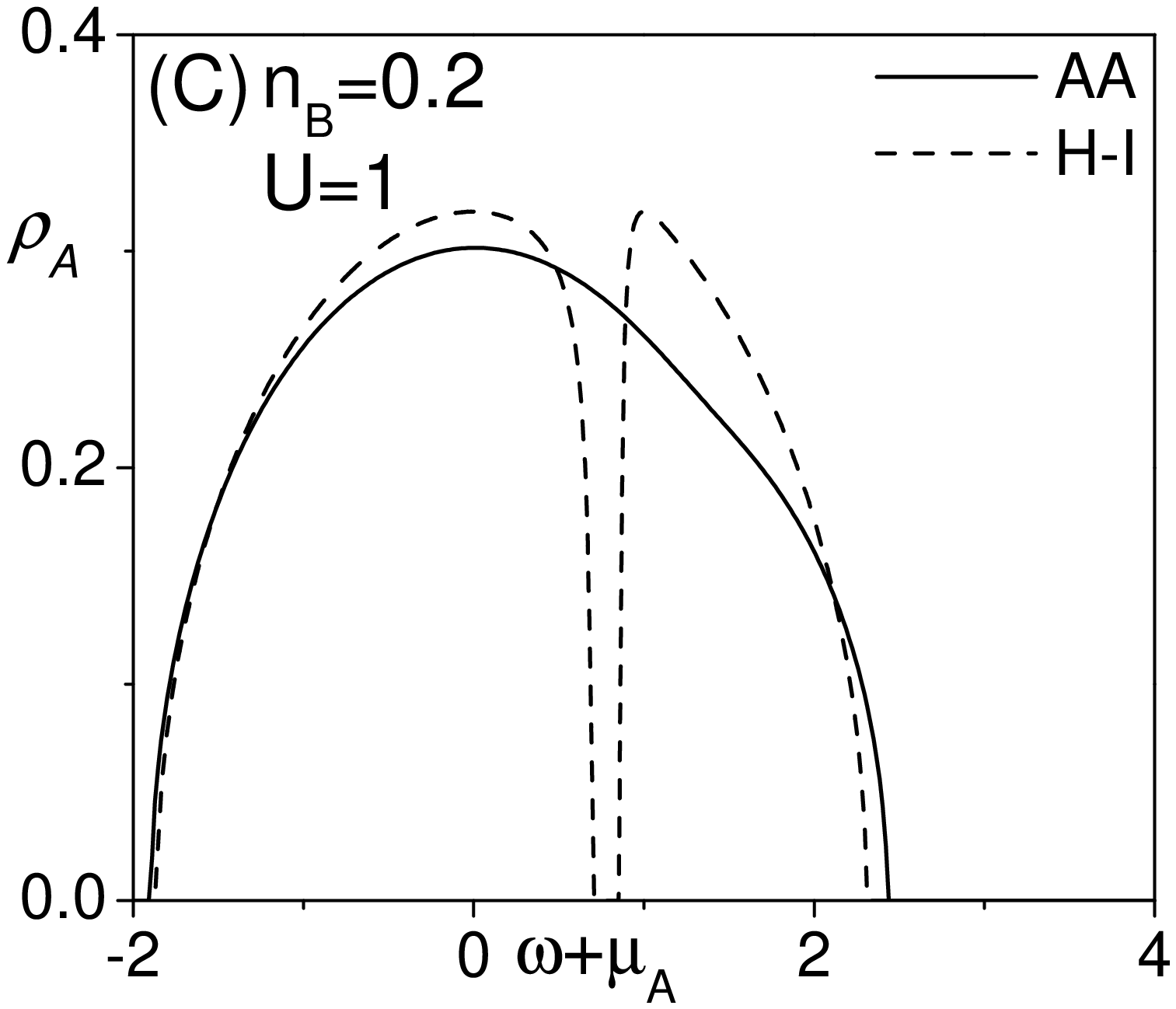}\hfill
\includegraphics[width=0.24\textwidth]{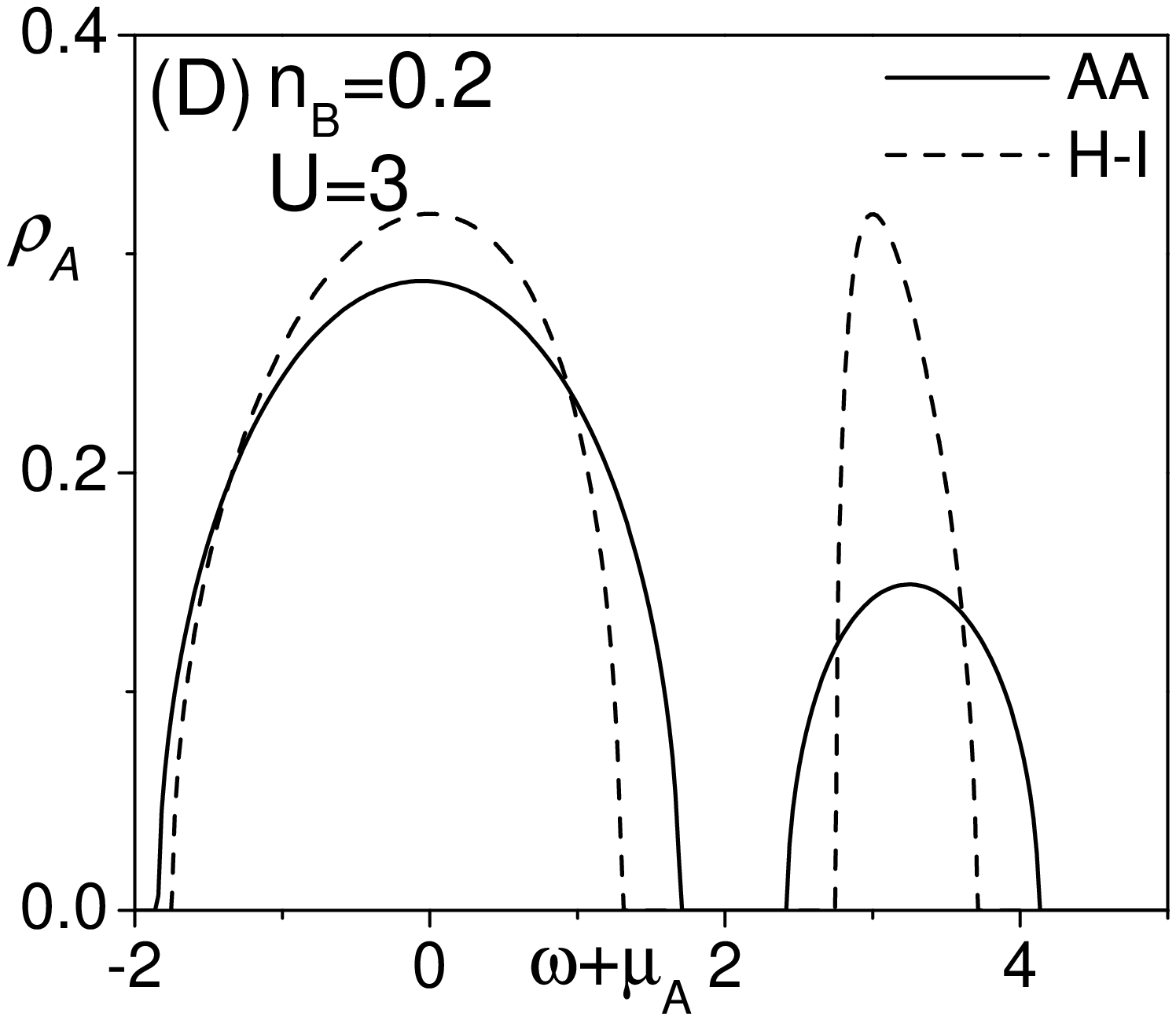}
{\caption{\label{fig_dos_aa_vs_hi} Density of states $\rho_A$ in
the alloy-analogy (solid line) and Hubbard-I (dashed line)
approximations on the Bethe lattice ($t_A=1$) for various $n_B$
and $U$.}}
\end{figure}

It is well known that the Hubbard-I approximation gives an
electron spectrum with the opened gap for any nonzero repulsion
$U$ (\fig\ref{fig_dos_aa_vs_hi}). Thus, this approximation cannot
describe metal-insulator transitions. However, we consider it here
to estimate its applicability at large $U$ from a thermodynamic
point of view.

The interacting single-site Green's function $G_{\sigma} (\omega)$
is calculated using the integration (\ref{G_from_Xi}) with a
noninteracting density of states $\rho^0_{\sigma}(\varepsilon)$.
The band of particles of a given type ($\sigma$) depends here only
on $\rho^0_{\sigma}(\varepsilon)$ and a concentration of particles
of the opposite type, i.e., the transfer parameter
$t_{\bar\sigma}$ does not have an effect on $G_{\sigma} (\omega)$.

In the large-$U$ limit ($U \gg t_\sigma$), the density of states
has the simple form
\begin{equation}
\rho_{\sigma}^{\mathrm{H-I}}(\omega) = \rho^0_{\sigma} \!\left(
\frac{\omega+\mu_\sigma}{1-n_{\bar\sigma}} \right)\! +
\rho^0_{\sigma} \!\left(
\frac{\omega+\mu_\sigma-U}{n_{\bar\sigma}} \right).
\end{equation}%
Both spectral subbands are of the same height for any
concentration values and are equal to
$\rho^0_{\sigma}(\varepsilon)$ with the scaled bandwidth.

\subsection{Alloy-analogy approximation}%

The system with one type of particles frozen on a lattice can be
mapped onto the problem of the electronic structure of simple
binary alloys \cite{Bra89,Fre90}. The alloy-analogy approximation
is the single-site solution of the binary alloy problem within the
coherent potential approximation \cite{Vel68} that is exact in
infinite dimensions. Such a mapping onto the binary alloy is exact
for the spinless Falicov-Kimball model with localized ions. Thus,
the alloy-analogy approximation exactly describes the band
formation by the particle hopping (transfer) and the interaction
with the localized particles. However, in the case of the
asymmetric Hubbard model this approximation does not include an
effect of the transfer of particles of another sort, like the
Hubbard-I approximation.

The self-energy part for the Dyson equation in the alloy-analogy
approximation reads
\begin{equation}
\Sigma_{\sigma}(\omega)=\frac{n_{\bar\sigma}
U}{1-G_{ii}^{\sigma}(\omega)(U-\Sigma_{\sigma}(\omega))}\,,
\end{equation}%
where the single-site Green's function:%
{\arraycolsep=0.4pt
\begin{eqnarray}
 G_{ii}^{\sigma}(\omega) & = & G_{\sigma}(\omega) \nonumber\\
      & = & \frac{1-n_{\bar\sigma}}
       {\omega+\mu_\sigma-J_{\sigma}(\omega)}
       + \frac{n_{\bar\sigma}}
       {\omega+\mu_\sigma-U-J_{\sigma}(\omega)}\,.
\end{eqnarray}}%
The Larkin irreducible part
$\Xi_{\sigma}^{-1}=\omega+\mu_\sigma-\Sigma_\sigma$ is determined
by the following expression
\begin{eqnarray}
 \Xi_\sigma^{-1}(\omega) & = & \bigg[ \frac{1-n_{\bar\sigma}}
       {\omega+\mu_\sigma-J_\sigma(\omega)}
       + \frac{n_{\bar\sigma}}
       {\omega+\mu_\sigma-U-J_\sigma(\omega)}
       \bigg]^{-1} \nonumber\\
& +&  J_\sigma(\omega).
\end{eqnarray}%
This approximation is obtained in our approach when the projecting
constant $\varphi_\sigma$ and the function $R_\sigma (\omega)$ in
(\ref{Omega}) are neglected.

\begin{figure}
\includegraphics[width=0.49\textwidth]{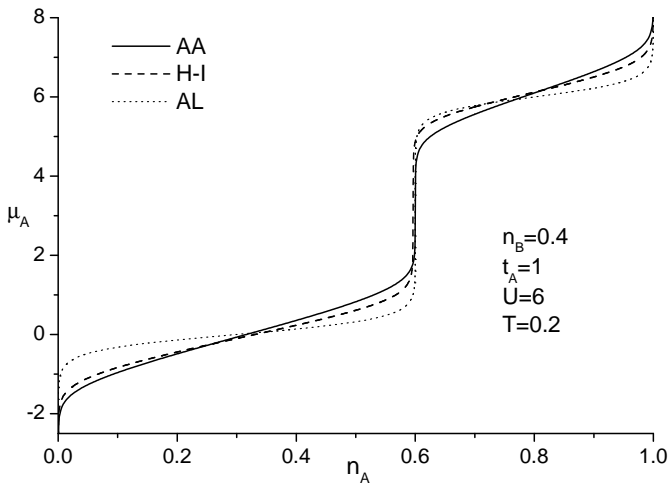}
\caption{\label{fig_mu_n} The chemical potential as a function of
the concentration in the alloy-analogy approximation (solid line)
on the Bethe lattice compared to the Hubbard-I approximation
(dashed line) and the atomic limit $t_\sigma \rightarrow 0$
(dotted line).}
\end{figure}

To compare analytically results of the alloy-analogy approximation
and the Hubbard-I approximation we can examine the equations in
the infinite-$U$ limit. The self-consistency condition gives
$J_{\sigma}=t_{\sigma}^2 G_{\sigma}$ on the Bethe lattice. Thus,
in this case, we have the following expression for the particle
spectrum:
\begin{equation}
\rho_{\sigma}^{\mathrm{AA}}(\omega)=\frac{1}{2 \pi t_{\sigma}^2}
\sqrt{4 t_{\sigma}^2 (1-n_{\bar\sigma})-(\omega+\mu_\sigma)^2}
\end{equation}%
when $|\omega+\mu_\sigma|<2 t_\sigma \sqrt{1-n_{\bar\sigma}}$ and
zero otherwise. The height and width of this spectral band depend
on the concentration $n_{\bar\sigma}$ and they are scaled
proportionally to $\sqrt{1-n_{\bar\sigma}}$ when $n_{\bar\sigma}$
changes. Therefore, the bandwidth of the subbands (for finite $U$)
is wider in the alloy-analogy approximation at half-filling
\fig\ref{fig_dos_aa_vs_hi}(A,C) as well as off half-filling
\fig\ref{fig_dos_aa_vs_hi}(B,D). In \fig\ref{fig_mu_n}, the effect
of wider subbands is illustrated by the dependence of the chemical
potential $\mu_A$ on the concentration $n_A$ calculated using the
particle density of states (\ref{rho_DOS}).

\section{Analytic properties at Mott transition}
\label{analytic_at_transition}

We consider here the metal-insulator transition at half-filling.
It is known that for the Hubbard model at zero temperature the
transition occurs when $\Imm \Sigma_{\sigma}(0)=0$ and the real
part has the following low-frequency form \cite{Geo96}
\begin{equation}
\Ree \Sigma_{\sigma}(\omega+ \ri 0^+) = U/2 +
(1-1/Z)\omega+\mathrm{O}(\omega^3)
\end{equation}
corresponding to Fermi liquid behaviour, where $Z$ defines the
quasiparticles weight. Such self-energy gives particle densities
of states with a quasiparticle peak  at the Fermi-level that has
the same height for different $U$ up to the critical value
$U_{\mathrm{c}}$ \cite{Bul99}. The gap opens only when
$Z\rightarrow 0$ and  in the metallic phase the relation
$G_{\sigma}(0) J_{\sigma}(0)=-1$ is satisfied.

However, for finite temperatures the Fermi-liquid state breaks
near the transition point $U_{\mathrm{c}}$. Therefore,
$\Sigma_{\sigma}(0)$ has a nonzero imaginary part \cite{Bul01},
and the limit $Z\rightarrow 0$ cannot indicate the transition. In
this case, the density of states at Fermi-level continuously tends
to zero. At high enough temperatures, the Mott transition turns
into a crossover from a bad metal to a bad insulator \cite{Roz94}.

Another limit of the asymmetric Hubbard model (the Falicov-Kimball
model) at half-filling does not have a central quasiparticle peak
in the spectrum \cite{Bra92,Fre05}. The alloy-analogy, Hubbard-III
and GH3 approximations also cannot describe the Fermi liquid
behaviour at half-filling. Hence, these approximations can be
applied to the investigation of high temperature properties of the
system when the quasiparticle features disappear.

Thus, to investigate the continuous transition in the asymmetric
Hubbard model, we derive some properties following from the
particle hole-symmetry.

\subsection{Exact relations between $G_\sigma (0)$, $J_\sigma (0) $ and $\Xi_\sigma(0)$
            at continuous crossover at half-filling}

\emph{In the case of the symmetric noninteracting density of
states
\begin{equation}
\rho^0_{\sigma}(\varepsilon)=\rho^0_{\sigma}(-\varepsilon)%
\label{rho_symmetry}
\end{equation}%
leading to the particle-hole symmetry with the properties
\begin{equation}
G_{\sigma}(\omega)=-G_{\sigma}(-\omega),\quad
J_{\sigma}(\omega)=-J_{\sigma}(-\omega)%
\label{hf_symmetry}
\end{equation}%
of the single-particle Green's function and the coherent potential
 at half-filling, the following relations can be proven
\begin{equation}
\lim_{U\rightarrow
U_{\mathrm{c}}^-}\frac{J_{\sigma}(0)}{G_{\sigma}(0)}=
\int_{-\infty}^{+\infty} \varepsilon^2 \rho^0_{\sigma}
(\varepsilon) \rd \varepsilon, \label{J_to_G_at_Mott}
\end{equation}%
\begin{equation}
\lim_{U\rightarrow U_{\mathrm{c}}^-} G_{\sigma}(0)
\Xi_{\sigma}^{-1}(0)= 1 \label{Xi_to_G_at_Mott}
\end{equation}%
which take place at zero frequency when the continuous
metal-insulator transition (crossover) is approached from below
(when the gap opens in the particle spectrum at $\omega=0$).
}\\[0ex]

\emph{Proof.} Let us consider general properties of the Green's
functions ($G_{\sigma}$, $J_{\sigma}$) and the total irreducible
part $\Xi_{\sigma}$ following from (\ref{hf_symmetry}) at zero
frequency. These functions can be written in the Lehmann
representation
\begin{equation}
F(\omega)=- \frac{1}{\pi}\lim_{\eta \rightarrow 0^+}
\int_{-\infty}^{+\infty}  \frac{\Imm F(\omega'+\ri
\eta)\rd\omega'}{\omega-\omega'}\,.
\end{equation}%
This representation shows that due to the symmetry
(\ref{hf_symmetry}) the functions at $\omega=0$ are pure
imaginary, and there are three possible cases: (i) $F(0)=0$; (ii)
$\Ree F(0)=0$, $\Imm F(0)=\const$; (iii) the pole of $F(\omega)$
in the real axis at ${\omega=0}$.

An imaginary part of $G_\sigma(\omega\pm0^+)$ defines the particle
spectrum (\ref{rho_DOS}). When the gap continuously opens the
single-particle Green's function tends to zero at ${\omega=0}$. It
means that $\Xi_{\sigma}^{-1}(0)$ is pure imaginary
\begin{equation}
\lim_{\eta \rightarrow 0^+} \Xi_{\sigma}^{-1}(\pm \ri \eta)=\pm
\ri B_{\sigma} , \quad B_{\sigma}>0,
\end{equation}%
and the gap opens (in the non-Fermi liquid state) only when
$B_{\sigma}$ continuously increases up to infinity, i.e., the
self-energy diverges \cite{Gro94,Geo96,Bul01}.

The coherent potential can be expressed from (\ref{G_Xi_J}) as a
function of $G_\sigma$ and the total irreducible part
$\Xi_\sigma$:
\begin{equation}
\frac{J_{\sigma}}{G_{\sigma}}=\frac{\Xi_{\sigma}^{-1}G_{\sigma}-1}{G_{\sigma}^2}\,.%
\label{eq_a_02}
\end{equation}%
Using that the noninteracting density of states is symmetric
(\ref{rho_symmetry}) and normalized
\begin{equation}
\int_{-\infty}^{+\infty} \rho^0_\sigma(\varepsilon) \rd
\varepsilon
= 1,%
\label{normalization}
\end{equation}%
we have:
\begin{equation}
G_{\sigma}= \mp \ri B_{\sigma} \int_{-\infty}^{+\infty}
\frac{\rho^0_\sigma (\varepsilon) \rd \varepsilon}{\varepsilon^2+ B_\sigma^2} \,,%
\label{eq_a_03}
\end{equation}%
and
\begin{equation}
G_\sigma J_\sigma= \Xi_{\sigma}^{-1}G_{\sigma}-1 = -
\int_{-\infty}^{+\infty} \frac{\varepsilon^2 \rho^0_\sigma
(\varepsilon) \rd \varepsilon}
{\varepsilon^2+ B_\sigma^2}%
\label{eq_a_04}
\end{equation}%
from (\ref{G_from_Xi}).

When the critical value $U_{\mathrm{c}}$ is approached from below,
the limit ${B_\sigma \rightarrow +\infty}$ should be considered.
In this limit, inserting (\ref{eq_a_03}) and (\ref{eq_a_04}) into
(\ref{eq_a_02}) leads to the result (\ref{J_to_G_at_Mott}).

To prove (\ref{Xi_to_G_at_Mott}), we consider three different
cases depending on behaviour of the function
$\rho^0_{\sigma}(\varepsilon)$ at large energies.
\begin{enumerate}
\item%
If an average kinetic energy per particle is finite, the limit
\begin{equation}
\lim_{\varepsilon \rightarrow \infty} \varepsilon^2
\rho^0_\sigma(\varepsilon) = 0
\end{equation}%
is satisfied giving the dependence $G_\sigma J_\sigma \sim
B_{\sigma}^{-2} $ (\ref{eq_a_04}) at large $B_{\sigma}$. It means
that $J_\sigma(0) \rightarrow 0$ at the transition.

\item%
When the density of states has the power-law tails of order
$\varepsilon^{-2}$
\begin{equation}
\lim_{\varepsilon \rightarrow \infty} \varepsilon^2
\rho^0_\sigma(\varepsilon) = \const > 0,
\end{equation}%
the limiting behaviour $G_\sigma J_\sigma \sim B_{\sigma}^{-1}$ is
realized at the transition. In this case a finite imaginary part
of $J_{\sigma}(0)$ remains when $B_{\sigma}\rightarrow\infty$, and
the ratio $J_{\sigma}(0)/G_{\sigma}(0)$ tends to infinity
(\ref{J_to_G_at_Mott}).

An example of such a function is the Lorentzian density of states
describing a lattice with long-range hopping along coordinate axes%
\begin{equation}
\rho^0_{\sigma} (\varepsilon) =
\frac{t_\sigma}{\pi(\varepsilon^2+t_{\sigma}^2)} \,.%
\label{Lorentzian_DOS}
\end{equation}
This density of states always gives a constant value for the
coherent potential:
\begin{equation}
\lim_{\eta \rightarrow 0^+} J_{\sigma}(\omega\pm \ri \eta) = \mp
\ri t_{\sigma}.
\end{equation}%

\item%
The last case is when $\rho^0_{\sigma}(\varepsilon) \sim
\varepsilon^{-n}$, $n \in (1,2)$ at large energy $\varepsilon$.
The integrals (\ref{normalization}) and (\ref{eq_a_04}) are
convergent only when $n>1$. Such power-law tails give the
following limiting behaviour $G_\sigma J_\sigma \sim
1/(B_{\sigma})^{n-1}$ and $J_\sigma \sim (B_{\sigma})^{2-n}$ at
large $B_{\sigma}$.
\end{enumerate}
All these cases show that the relation
\begin{equation}
\lim_{U\rightarrow U_{\mathrm{c}}^-} G_{\sigma}(0) J_{\sigma}(0)=
-0
\end{equation}%
is satisfied proving the relation (\ref{Xi_to_G_at_Mott}).

\subsection{Critical value $U_{\mathrm{c}}$ for the
Mott transition within GH3 approximation  with symmetric density
of states}

Due to (\ref{J_to_G_at_Mott}), we have
\begin{equation}
J_{\sigma}(0)=W_{\sigma}^2 G_{\sigma}(0)/4%
\label{JGW2}
\end{equation}%
at the Mott transition, where $W_{\sigma}$ is the effective
half-bandwidth of the unperturbed density of states
$\rho^0_\sigma$:
\begin{equation}
W_{\sigma}=2 \left( \int_{-\infty}^{+\infty} \varepsilon^2
\rho^0_{\sigma} (\varepsilon) \rd \varepsilon \right)^{1/2} .
\end{equation}%

Due to $G_{\sigma}(0) \Xi_{\sigma}^{-1}(0)=1$
(\ref{Xi_to_G_at_Mott}) at the transition, inserting (\ref{JGW2})
into (\ref{Xi_halffilling}) yields a set of equations:
\begin{equation}
\lim_{U\rightarrow U_{\mathrm{c}}^-} \frac{U^2 G_{A}(0)}{W_A^2
G_{A}(0) + 2 W_B^2  G_{B}(0)}=1 \,,
\end{equation}%
\begin{equation}
\lim_{U\rightarrow U_{\mathrm{c}}^-} \frac{U^2 G_{B}(0)}{W_B^2
G_{B}(0) + 2 W_A^2  G_{A}(0)}=1 \,.
\end{equation}%
The ratio
\begin{equation}
\eta=\lim_{U\rightarrow U_{\mathrm{c}}^-}
\frac{G_{A}(0)}{G_{B}(0)}
\end{equation}%
is excluded from the set of equations, and we have the following
equation for $U_{\mathrm{c}}$
\begin{equation}
U_{\mathrm{c}}^4-(W_A^2+W_B^2) U_{\mathrm{c}}^2 -3 W_A^2 W_B^2=0\,
\end{equation}%
with the solution
\begin{equation}
U_{\mathrm{c}}= \sqrt{\frac{W_A^2+W_B^2+(W_A^4+W_B^4+14 W_A^2
W_B^2)^{1/2} }{2}} \,. \label{Uc_hf_ahm}
\end{equation}%

If we put $W_B=0$, the expression (\ref{Uc_hf_ahm}) describes the
Mott-type transition in the alloy-analogy approximation (which is
exact for the Falicov-Kimball model) giving well known results:
$U_{\mathrm{c}}=W_A=2 t_A$ for the Bethe lattice and
$U_{\mathrm{c}}= \sqrt{2} t_A$ for the hypercubic lattice. In the
case of the standard Hubbard model ($W_A=W_B=W$), we have the
result $U_{\mathrm{c}}= \sqrt{3} W$ corresponding to the
Hubbard-III approximation at half-filling (see
Refs.~\onlinecite{Fre03,Geo96} for reviews of the limits: the
standard Hubbard model and the Falicov-Kimball model).

\section{Results and discussion}
\label{result_discussion}

There are two different pairs of parameters ($n_A$, $\mu_A$ and
$n_B$, $\mu_B$) in the asymmetric Hubbard model which have to be
determined from the thermodynamic equilibrium. Thus, various
thermodynamic regimes may be considered when different pairs of
the thermodynamic parameters are fixed: ($\mu_A$, $\mu_B$),
($\mu_A$, $n_B$), ($n_A$, $n_B$), etc. It is known that there are
possible phase separations at low temperatures in the case of the
Falicov-Kimball model \cite{Fre99,Let99,Sta03} and in the general
case of the asymmetric Hubbard model \cite{Uel04}. Besides the
segregated phases, the long-range antiferromagnetic-type ordering
is possible at low temperatures \cite{Koc94}.

Since the investigation of the asymmetric Hubbard model is a very
complicated problem, we restrict our analysis to temperatures
higher than the critical temperatures of thermodynamic
instabilities. Thus, the problem is reduced to the investigation
of a band structure and metal-insulator transitions at constant
particle concentrations.

The simplest approximations such as the Hub\-bard-I and
alloy-analogy approximations describe the band of particles
generated by the particle hopping and the interaction with
particles of another sort. These approximations can describe an
effect of the transfer parameter $t_A$ on the spectrum $\rho_{B}
(\omega)$ only via the concentration $n_A$, i.e., when the
concentrations are fixed the band $\rho_{B} (\omega)$ does not
depend on $t_A$, and for $t_A\neq 0$, $t_B=0$ the approximate
spectrum $\rho_B (\omega)$ has a form of two delta-peaks. However,
the alloy-analogy approximation is exact only for itinerant
particles in the Falicov-Kimball model when another sort of
particles is localized. Thus, this approximation  can give simple
reasonable results for dependence $\mu_\sigma=\mu_\sigma
(n_\sigma, n_{\bar\sigma})$ in the limit of small values of
$n_{\bar\sigma}$ or $t_{\bar\sigma}$.

The generalization of the Hubbard-III approximation (GH3) gives
the total irreducible part and the single-site Green's function as
a functional of the coherent potentials of both sorts of particles
(see equations (\ref{irreduciblepart})-(\ref{R_omega})). Unlike
the alloy-analogy approximation, GH3 described broadening of the
band by the interaction with particles of another sort, and it
gives a finite bandwidth for localized particles in the
Falicov-Kimball model. In general, the spectral density in the GH3
approximation is temperature dependent; such an approximation can
be applied for systems with different values of the transfer
parameters (both $t_A=t_B$ and $t_A\neq t_B$) and various particle
concentrations. However, the approximation has some restrictions.
In Ref.~\onlinecite{Sta05}, GH3 was applied to the calculation of
the chemical potential and spectrum of localized particles in the
Falicov-Kimball model. It was shown that the approximation gives
better results when the concentration of itinerant particles is
low.

\begin{figure}
\begin{center}
\hspace*{-5ex}\includegraphics[width=0.4\textwidth]{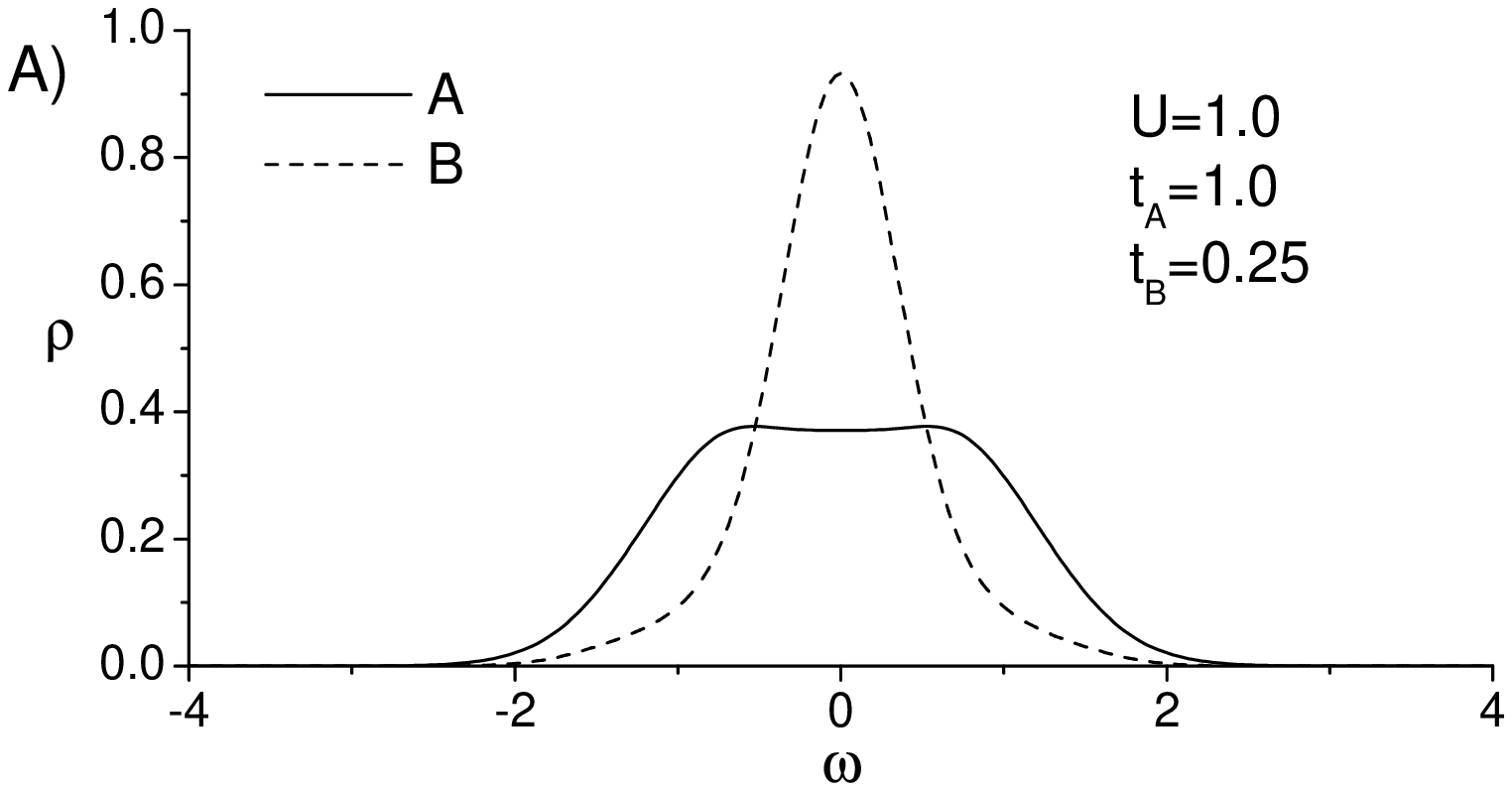}\\[-2.0ex]
\hspace*{-5ex}\includegraphics[width=0.4\textwidth]{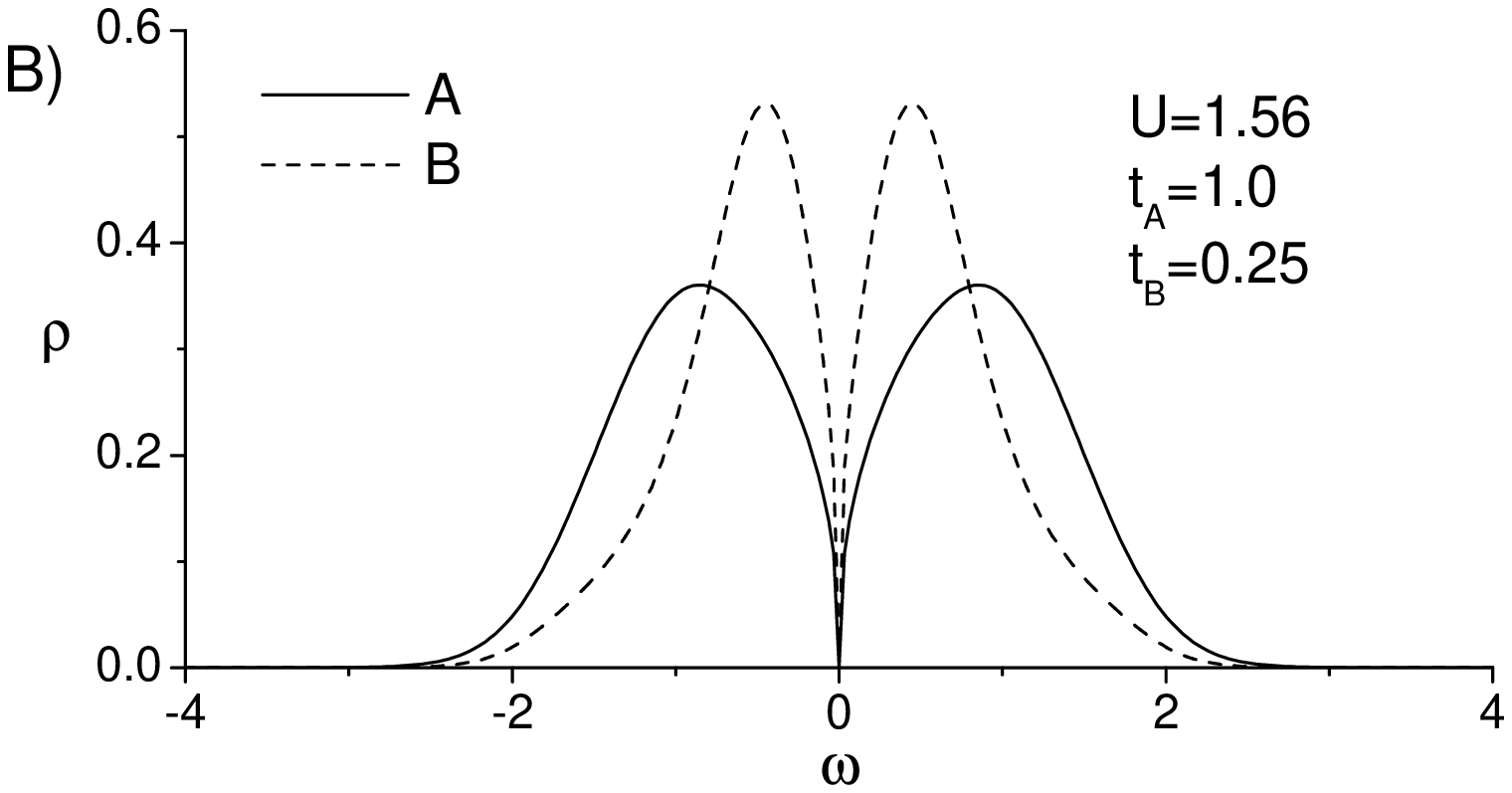}\\[-2.0ex]
\hspace*{-5ex}\includegraphics[width=0.4\textwidth]{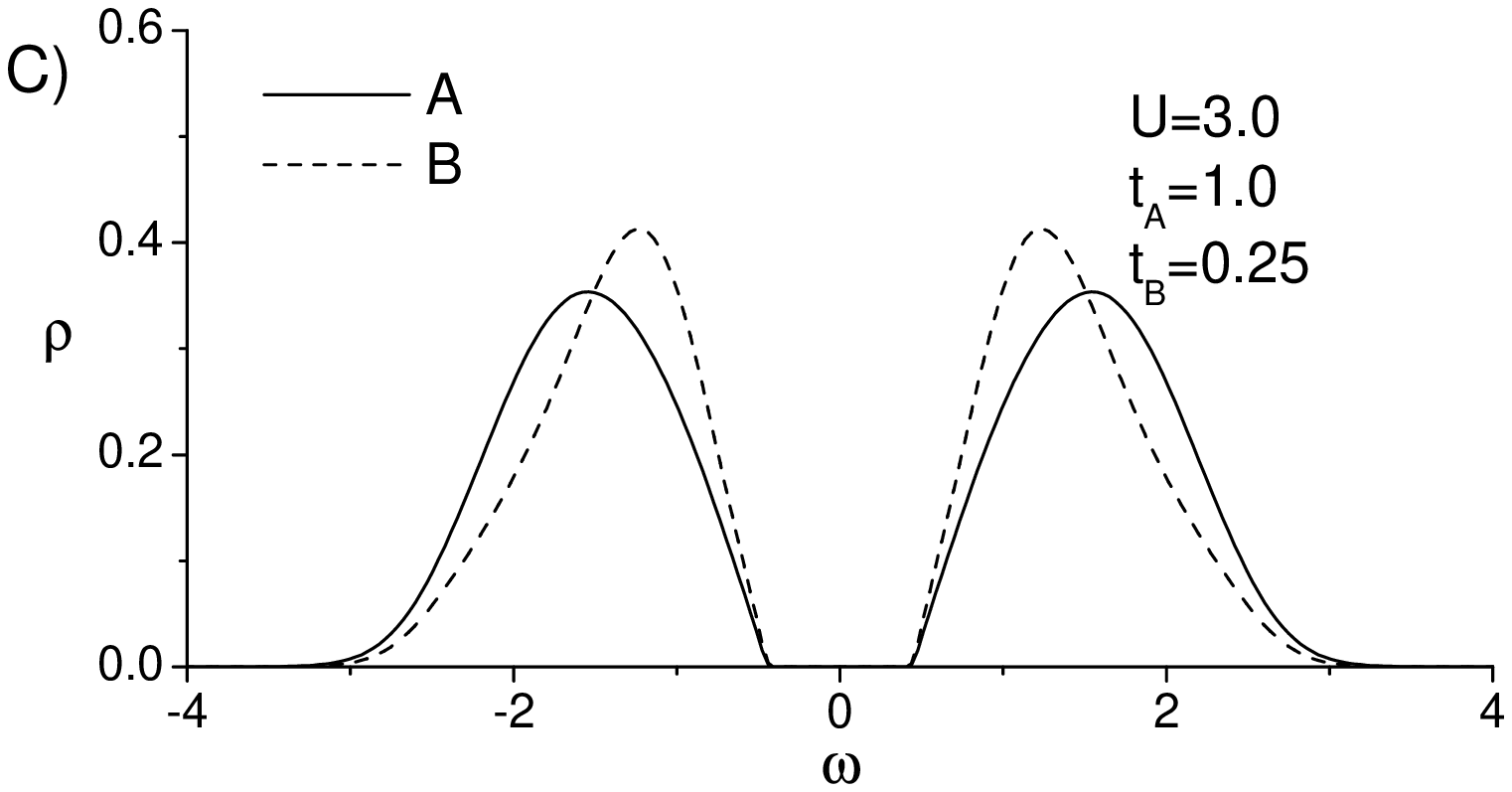}
\end{center}
\caption{\label{fig_dos_mott} Densities of states of particles in
the asymmetric Hubbard model within the GH3 approximation for
various values of $U$ at half-filling on the hypercubic lattice.}
\end{figure}

In the symmetric case of half-filling the approximate density of
states is independent of temperature (\fig\ref{fig_dos_mott}),
because the projecting coefficients and the coefficients of the
integral terms in $R_\sigma (\omega)$  (\ref{R_omega}) are equal
to zero. In the Falicov-Kimball model at half-filling, the
approximation (\ref{hf_FK}) can be compared with the exact results
\cite{Bra92,Fre05}. We find that the approximation gives a better
correspondence with the exact curves at high temperatures, and the
best correspondence is for high $U$ when the exact results weakly
depend on temperature \cite{Sta05}. \fig\ref{fig_dos_mott}(A)
shows that the density of states of heavy particles B (the case of
$t_B<t_A$) is in the form of a single peak for the low interaction
$U$. For the higher interaction strength
\fig\ref{fig_dos_mott}(B,C), the spectrum $\rho_B(\omega)$ has two
subbands with peaks which are closer to the centre ($\omega=0$)
than in the spectrum of light particles $\rho_A(\omega)$. This
agrees with the results obtained for the Falicov-Kimball model
when $t_B=0$.

The particle-hole symmetry at half-filling simplifies the
investigation of the problem. Such symmetry requires the
divergence ($1/\omega$) of the self-energy at zero frequency when
the gap opens continuously \cite{Gro94,Bul01}. Thus, the limit
behaviour (\ref{J_to_G_at_Mott}), (\ref{Xi_to_G_at_Mott}) of the
Green's function and the coherent potential at the transition is
proven analytically, which allows us to perform some analytic
analysis.

In \fig\ref{fig_dos_mott}, the transition with the opening gap is
illustrated for the asymmetric Hubbard model within the GH3
approximation. There are only continuous transitions. This agrees
with the fact that the Mott transition has to be of a continuous
type at high temperatures for the standard Hubbard model
\cite{Sch99,Bul01}, and in the case of the Falicov-Kimball model
the gap opens continuously with increasing $U$ \cite{Fre03}. The
spectra of both types of particles have the same bandwidth in the
approximation. This is because it cannot describe tails of the
subbands that raise with the temperature increase as it is for
localized particles in the Falicov-Kimball model
\cite{Bra92,Fre05}. Since the effect of temperature broadening of
the band is not captured, it can be predicted that the
approximation underestimate the critical value $U_{\mathrm{c}}$.

\begin{figure}
\includegraphics[width=0.47\textwidth]{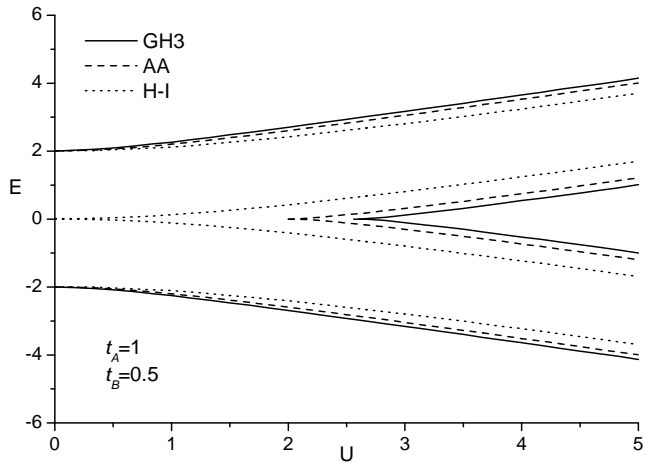}
\caption{\label{fig_band_U} Dependence of the energy band edges on
the repulsion strength $U$ at half-filling $n_A=n_B=1/2$ on the
Bethe lattice: ${t_A=1}$, ${t_B=0.5}$. The solid line is the GH3
approximation, the dashed and dotted lines are the alloy-analogy
and Hubbard-I approximations, respectively.}
\end{figure}

\begin{figure}
\includegraphics[width=0.49\textwidth]{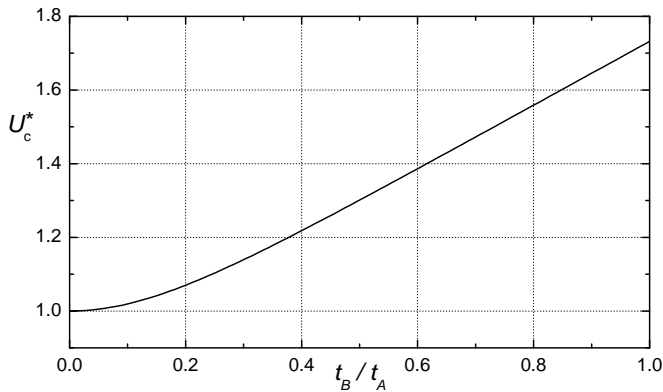}
\caption{\label{fig_u_tt} Scaled critical value
$U^{*}_{\mathrm{c}}=U_{\mathrm{c}}/W_A$ (\ref{Uc_scaled}) as a
function of $t_B/t_A$ within the GH3 approximation at
half-filling.}
\end{figure}

The energy band edges as a function of $U$ within various
approximations on the Bethe lattice are displayed in
\fig\ref{fig_band_U}. In the Hubbard-I approximation the spectrum
is always split having the gap and two subbands. The alloy-analogy
approximation is exact for the band of itinerant particles when
$t_B=0$ and it gives the critical interaction constant
${U^{\mathrm{AA}}_{\mathrm{c}}=2t_A}$. The GH3 approximation
describes the band which is broadened by simultaneous contribution
of the hopping of both types of particles, and the critical value
$U_{\mathrm{c}}>U^{\mathrm{AA}}_{\mathrm{c}}$. Because of the
particle-hole symmetry at half-filling the critical value
$U_{\mathrm{c}}$ can be calculated analytically in the GH3
approximation and it is given by the expression (\ref{Uc_hf_ahm}),
see \fig\ref{fig_u_tt}. The obtained result can be applied to
various lattices (hypercubic, Bethe, etc.) when the energy is
scaled in the following way:
\begin{equation}
U_{\mathrm{c}}^{*}= \frac{U_{\mathrm{c}}}{W_A} =
\frac{U_{\mathrm{c}}}{2} \left( \int_{-\infty}^{+\infty}
\varepsilon^2 \rho^0_A (\varepsilon) \rd \varepsilon
\right)^{-1/2}
.%
\label{Uc_scaled}
\end{equation}%
It is common practice to normalize $U$ by a typical kinetic energy
($W_A$) investigating the Mott transition in the standard Hubbard
model \cite{Geo96}. In this case, such normalization shows that
the scaled result does not depend on a form of the noninteracting
density of states. For the Lorentzian density of states
(\ref{Lorentzian_DOS}) with long-range hopping an average kinetic
energy  per particle tends to infinity ($W_A\rightarrow\infty$),
and as it was noted in Ref.~\onlinecite{Geo92b}, this requires an
infinite interaction $U$ to drive the system insulating.

\section{Conclusions}
\label{conclusions}

The generalization of the Hub\-bard-III approximation is obtained
for the asymmetric Hubbard model using the equation of motion
approach. This method combines in a unified framework the
description of the band structure of the Falicov-Kimball model and
the standard Hubbard model. In general, the approach can be used
to calculate the particle spectrum in the system with the
different particle concentrations at various temperatures, and it
gives nontrivial results for the spectrum of localized particles
in the Falicov-Kimball model.

The self-consistency condition connecting mutually the single-site
Green's function and the coherent potential gives the simple
result (\ref{J_to_G_at_Mott}) when the metal-insulator transition
occurs at half-filling on the lattice with a symmetric
noninteracting density of states. As a result, we have the
expression (\ref{Uc_hf_ahm}) in the GH3 approximation for the
critical value of $U$ that provides a universal solution for
various lattices and different values of the transfer parameters.

The approximation has some restrictions. Thus, to describe the
system at half-filling with the proper temperature dependence or
the quasi particle peak, the approach needs further improvement in
calculating the mass operator by including higher order
corrections.

\end{document}